\documentclass[12pt]{article}
\InputIfFileExists{epsf.sty}{}{}
\usepackage{graphicx}
\def\bpipi{B \to \pi \pi}
\def\bropi{B \to \rho \pi}
\def\beq{\begin{equation}}
\def\eeq{\end{equation}}
\def\bea{\begin{eqnarray}}
\def\eea{\end{eqnarray}}

\def\apz{A^{+0}}

\begin{document}

\begin{flushright}
UdeM-GPP-TH-04-125 \\
\end{flushright}

\begin{center}
\bigskip
{\Large \bf New Physics Signals through CP Violation in $B \to \rho \pi$ \footnote{talk given at {\it MRST 2004: From Quarks to
Cosmology}, Concordia University, Montreal, May 2004.}} \\

\bigskip
\bigskip

{\large Veronique Page \footnote{veronique.page@umontreal.ca}}
\end{center}

\begin{center}
{\it Laboratoire Ren\'e J.-A. L\'evesque, Universit\'e de Montr\'eal,}\\
{\it C.P. 6128, succ. centre-ville, Montr\'eal, QC,
Canada H3C 3J7} 
\end{center}

\begin{center} 
\bigskip (\today)
\vskip0.5cm
{\Large Abstract\\}
\vskip3truemm

\parbox[t]{\textwidth} {We describe here a method for detecting physics beyond the
standard model via CP violation in $B \to \rho \pi$ decays. Using a
Dalitz-plot analysis to obtain $\alpha$, along with an analytical
extraction of the various tree ($T$) and penguin ($P$) amplitudes, we
obtain a criterion for the absence of new physics (NP). This
criterion involves the comparison of the measured $|P/T|$ ratio with
its value as predicted by QCD factorization. We show that the
detection of NP via this method has a good efficiency when compared
with the corresponding technique using $B \to \pi \pi$ decays.}
\end{center}

\thispagestyle{empty}
\newpage
\setcounter{page}{1}
\baselineskip=14pt

\section{Introduction}

\begin{figure}[h]
\begin{center}
\includegraphics[angle=270, width=0.65\textwidth]{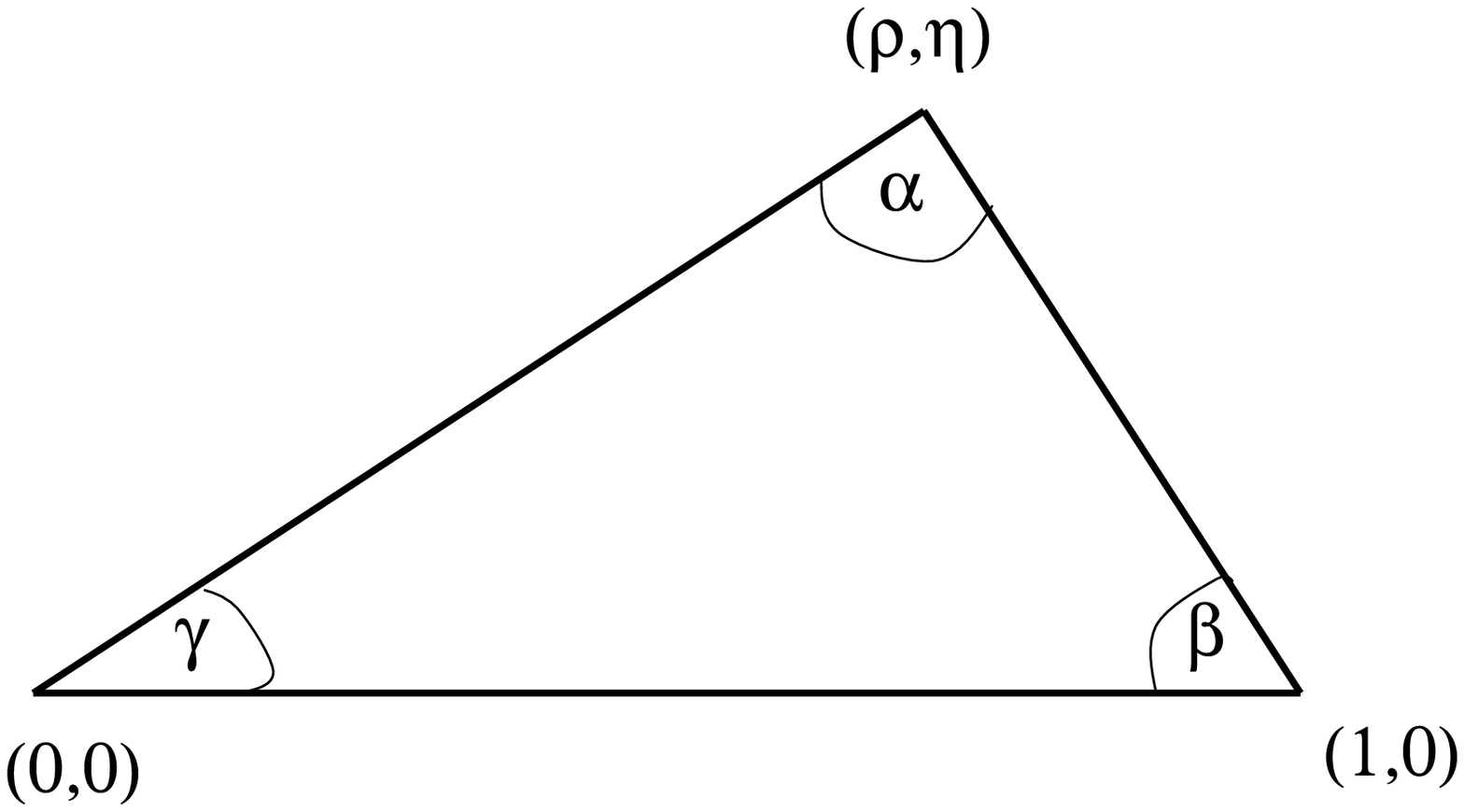}	
\caption{Unitarity triangle, with weak phases definitions}
\label{fig:unitaritytriangle}
\end{center}
\end{figure}

In the Standard Model (SM), CP violation is due to a non-trivial
complex phase in the quark mixing matrix, the CKM matrix. This phase
is most often described using the unitarity triangle
(Fig.~\ref{fig:unitaritytriangle}). Here $\alpha$, $\beta$ and
$\gamma$ are the three weak phases that can cause amplitudes
contributing to a given decay to interfere in such a way as to create
CP violation in this decay \cite{londonmrst}.

There have been many ways suggested to measure these three phases, the
ultimate goal being to overconstrain the unitarity triangle in order
to test the validity of the SM description of CP violation. Another
way to try and test the SM is to look for a discrepancy between the SM
predictions of CP violation in a given decay and the observed
values. The method described here is of the latter type. We will
review in the next section how $B \to \pi \pi$ has been used to date
for both measuring $\alpha$ \cite{isospin,charles} and detecting
physics beyond the SM \cite{lss01}. Since $B \to \pi \pi$ is not
dominated by one amplitude, but involves both tree ($T$) and penguin
($P$) contributions (a situation refered to as {\it penguin
pollution}), the indirect CP asymmetry does not lead directly to the
determination of $\alpha$. However, by using an isospin analysis of
$B\to\pi\pi$ decays, \cite{isospin}, one can extract $\alpha$ through
geometrical means (up to discrete ambiguities). Moreover, new physics
(NP) detection is shown to be possible despite the so-called {\it CKM
ambiguity} \cite{lss99}, but the technique loses some of its
efficiency due to the discrete ambiguities on $\alpha$.

The subsequent section will explain how $B \to \rho \pi$ can itself be
used for NP detection through the same method used in $\bpipi$.
Interestingly enough, a Dalitz plot analysis of $B \to \pi\pi\pi$
leads to an unambiguous determination of $\alpha$. Thus, though quite
challenging from the experimental point of view, $B \to \rho \pi$ is
shown to be very efficient in the detection of NP.

\section{$B \to \pi \pi$}

\subsection{$\bpipi$ without New Physics}

\begin{figure}
\begin{center}
\includegraphics[angle=270, width=0.65\textwidth]{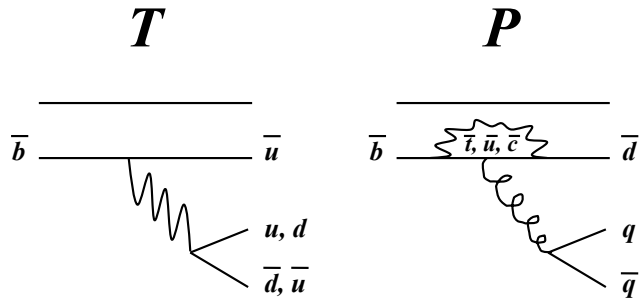}	
\caption{Tree and penguin amplitudes}
\label{fig:tetp}
\end{center}
\end{figure}

The $\bpipi$ system includes three decays (and three CP-conjugate
decays): two neutral decays, $B^0 \to \pi^+ \pi^-$ and $B^0 \to \pi^0
\pi^0$, and one charged, $B \to \pi^+ \pi^0$. In general, the decays
involve a $T$ and a $P$ contribution (see Fig.~\ref{fig:tetp}), so
that all decay amplitudes take the form
     \beq
     A^i = T^i e^{-i\alpha} + P^i ~,
     \eeq

\begin{figure}
\begin{center}
\includegraphics[angle=270, width=0.65\textwidth]{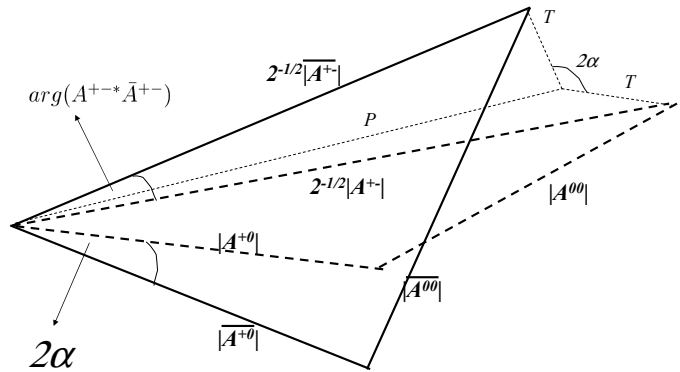}	
\caption{Isospin triangles.  $2\alpha$ appears at two places; one (between the $T$ diagrams of $A^{+-}$ and $\bar A^{+-}$) would be difficult to obtain, while the other one is obtained through simple geometrical means.}
\label{fig:isospintri}
\end{center}
\end{figure}

with (i=+-,00,+0). (The amplitudes have all been rescaled by
$e^{i\beta}$ so as to remove the mixing phase, and $\apz$ has no $P$
contribution.) Although one can measure indirect CP violation (${\rm
Im} \lambda$) in the two neutral decays,
     \beq  
     {\rm Im} \lambda = \sin {\bar A \over A} = \sin \left({
     Te^{i\alpha} + P \over Te^{-i\alpha} +P} \right) ~,
     \eeq
these measurements {\it cannot} be related to $\alpha$ in a simple
way. This is where isospin analysis \cite{isospin} enters the scene.
Decomposing all amplitudes and contributions in term of their isospin
content, one obtains the two following {\it triangle relationships}:
     \bea
     \frac{1}{\sqrt{2}} A^{+-} +A^{00} &=& A^{+0} ~, \nonumber \\ [4pt]
     \frac{1}{\sqrt{2}} \bar A^{+-} + \bar A^{00} &=& \bar A^{+0} ~.
     \eea
That is, one sees that $A^{+-}$, $A^{00}$ and $A^{+0}$ form a triangle
in isospin space (as do their CP-conjugate counterparts). $2\alpha$ is
then obtained through simple geometrical means. However, this
resolution will give $2\alpha$ up to an eightfold discrete ambiguity
(see Fig.~\ref{fig:isospintri}).

\subsection{$\bpipi$ with New Physics}

Let us now try and test the sensitivity of $\bpipi$ to New Physics
(NP). It is expected that NP will affect principally flavor-changing
neutral current (FCNC) diagrams \cite{np1,np2}. Thus both the box
diagram of the $B^0$ -- $\bar B^0$ mixing and the penguin diagram
could be affected, each (in general) in a different way. This would in
turn introduce a phase discrepancy between them. The amplitudes would
now appear as:
     \beq
     A^i = T^i e^{-i\alpha} + P^i e^{-i\theta_{NP}} ~.
     \eeq
Unfortunately, it has been shown \cite{lss99} that it is fundamentally
impossible to detect a penguin phase, due to the so-called {\it CKM
ambiguity}. To remove this ambiguity, it is necessary to make an
assumption about some hadronic parameters (or some combination of
them). One can then detect the presence of $\theta_{NP}$ by comparing
the theoretical SM prediction of this hadronic parameter with its
measured value. This analysis was done recently \cite{lss01}. The
criterion obtained is as follows: if $\theta_{NP} = 0$, the
observables respect
     \beq
     0.05 \le \frac{1-\sqrt{1-(a_{dir}^{+-})^2}
       \cos(2\alpha-2\alpha_{eff})}{1-\sqrt{1-(a_{dir}^{+-})^2}
       \cos(2\alpha_{eff})} \le 0.5 ~,
     \eeq
where $a_{dir}^{+-}$ and $2\alpha_{eff}$ are the direct and indirect
CP asymmetries, respectively. If it is found experimentally that this
inequality is not respected, it would indicate that $\theta_{NP} \ne
0$. The main experimental difficulty here is the large number of
discrete ambiguities in the extraction of $\alpha$ from $B\to\pi\pi$
decays. NP detection with this channel would thus probably necessitate
using $\alpha$ as obtained independently in some other decay.

\section{$B \to \rho \pi$}

$\bropi$ decays offer on this subject an interesting alternative. The
$\bropi$ system contains five decays (and five CP-conjugate decays):
three neutral decays ($B^0 \to \rho^+ \pi^-$, $B^0 \to \rho^- \pi^+$
and $B^0 \to \rho^0 \pi^0$) and two charged ($B^+ \to \rho^+ \pi^0$
and $B^+ \to \rho^0 \pi^+$). All amplitudes receive both a tree and a
penguin contribution:
     \beq
     S^i = T^i e^{-i\alpha} + P^i ~, 
     \eeq
($i=+-,-+,00,+0,0+$ where the first superscript is for $\rho$ and the
second is for $\pi$). Again all amplitudes have been rescaled by
$e^{i\beta}$. In this case too, then, we have penguin pollution.
Although an isospin analysis would give a pentagon relationship,
enabling one to solve geometrically for $2\alpha$, this resolution
would be painfully plagued by discrete ambiguities. In this case, as
explained in Ref.~\cite{dalitz}, it is the Dalitz-plot analysis of
$B^0 \to \pi^+ \pi^- \pi^0$ that saves the day. Since all three
neutral amplitudes contribute to $B^0 \to \pi^+ \pi^- \pi^0$ (and all
three neutral CP-conjugate amplitudes to $\bar B^0 \to \pi^- \pi^+
\pi^0$), one can use the interference between them to obtain $2\alpha$
unambiguously.

\subsection{$\bropi$ without New Physics}

In the SM case, the Dalitz plot contains in fact enough information to
ensure that one can solve for all theoretical parameters. We define
for each decay a branching ratio:
     \beq
     B_i = \frac{1}{2}(|S_i|^2 + |\bar S_i|^2)
     \eeq
and a direct CP asymmetry:
     \beq
     a_i = \frac{|S_i|^2 - |\bar S_i|^2}{|S_i|^2 + |\bar S_i|^2} ~.
     \eeq
We also define a indirect CP asymetry for the neutral decays:
     \beq
     2\alpha_{eff}^i = \mathbf{Arg}(\bar S_iS_i^*) ~.
     \eeq
The Dalitz-plot analysis ensures that all theoretical parameters can
in principle be expressed in term of these observables. In particular,
the penguin and tree amplitudes are solvable analytically. The results
we obtain are as follows: for each decay, the ratio of the penguin to
the tree amplitude is \cite{ropi}:
     \beq \label{eq:ratio}
     r^i \equiv \left\vert \frac{P^i}{T^i} \right\vert =
     \sqrt{\frac{1-\sqrt{1-a_i^2}\cos(2\alpha_{eff}^i-2\alpha)}
     {1-\sqrt{1-a_i^2}\cos2\alpha_{eff}^i}} ~.
     \eeq
This expression is the same for the two charged decays, with
$2\alpha_{eff}^i$ put to zero. The ratio is expressed in terms not
only of observables, but also of $2\alpha$; one can simply use for
this the value obtained through the Dalitz-plot analysis of the
$B\to\rho\pi$ system.

\subsection{$\bropi$ with New Physics}

Let us now add new physics in the FCNC, just as in the $\bpipi$
channel. The amplitudes are modified to:
     \beq
     S^i = T^i e^{-i\alpha} + P^ie^{-i\theta_{NP}} ~.
     \eeq
Just as in the case of $\bpipi$, the CKM ambiguity causes
$\theta_{NP}$ to be impossible to measure. However, here too we can
remove the CKM ambiguity by making an assumption about hadronic
parameters. We are still able to solve for each penguin-to-tree ratio
analytically:
     \beq \label{eq:ratioNP}
     r^i = \left\vert \frac{P^i}{T^i} \right\vert =
     \sqrt{\frac{1-\sqrt{1-a_i^2}\cos(2\alpha_{eff}^i-2\alpha)}
     {1-\sqrt{1-a_i^2}\cos(2\alpha_{eff}^i - 2\theta_{NP})}} ~.
     \eeq

One question one can ask is whether there are hidden contributions of
$\theta_{NP}$ in these expressions. Since up to now we've used the
Dalitz plot to obtain $2\alpha$, we have to make sure that this
analysis still allows its determination if $\theta_{NP} \ne 0$. The
point is that all penguins have been affected in the same way, so that
the interference between the neutral decays that allowed the
extraction of $2\alpha$ is left unchanged. Thus, with or without NP,
the Dalitz-plot analysis of $B \to \pi^- \pi^+ \pi^0$ gives $2\alpha$
unambiguously.

Eq.~(\ref{eq:ratioNP}) then stands as a testing ground for the
presence of new physics. Given the SM prediction of one (or some) of
the ratios, we could compare it with its measured value [as given by
Eq.~(\ref{eq:ratio})] and decide whether it is compatible with
$\theta_{NP} = 0$. 

A SM computation of $r^{+-}$ and $r^{-+}$ has recently been made
within the framework of QCD factorization \cite{qcdf}. The results are
as follows:
     \bea
     r^{+-} &=& 0.10^{+0.06}_{-0.04} ~, \nonumber \\[4pt]
     r^{-+} &=& 0.10^{+0.09}_{-0.05} ~.
     \eea
However, as we want to test for the presence of NP and not the
accuracy of QCD factorization (or, for instance, our assumption of no
electroweak penguins), we must be as conservative as possible
regarding the ranges used to compare with data. The above ranges are
therefore enlarged to include potential underestimate of errors. Our
criterion for the absence of NP will then be:
     \beq \label{eq:range}
     0.05 < r^i < 0.25 ~.
     \eeq
or, in terms of observables,
     \beq \label{eq:criterion}
     0.05 < \sqrt{\frac{1-\sqrt{1-a_i^2}\cos(2\alpha_{eff}^i-2\alpha)}
       {1-\sqrt{1-a_i^2}\cos2\alpha_{eff}^i}} < 0.25
     \eeq
for $i=+-,-+$.

\subsection{Sensitivity to NP}

The last question we want to consider is how restrictive our
criterion is. If it happened, for instance, that for every fixed value
of $2\alpha$, there existed for most values of $a_i$ a value of
$2\alpha_{eff}^i$ such that Eq.~(\ref{eq:criterion}) is satisfied,
then our criterion would effectively be useless. We must therefore
test the restrictiveness of Eq.~(\ref{eq:criterion}). 

\begin{figure}
\centerline{\epsfxsize 4.5 truein \epsfbox {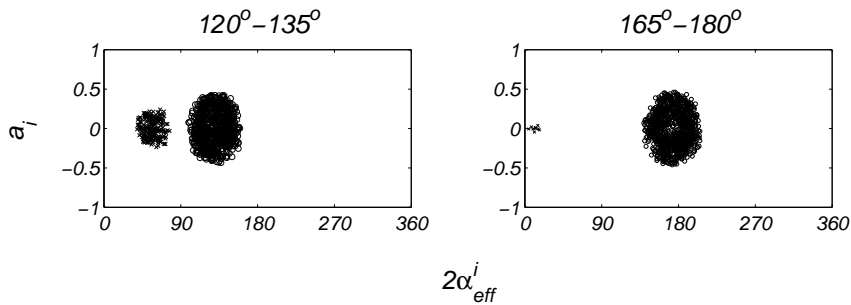}}
\caption{Regions in  $(2\alpha_{eff}^i,a_i)$ $(i = +-,-+)$ parameter space consistent with the actual (conservative) QCD factorization prediction on  $|P^i/T^i|$. $2\alpha$ is assumed to be known up to a 2-fold ambiguity; should it be obtained unambiguously, only the left-hand region would remain. \label{fig:results}}
\end{figure}

We first suppose $2\alpha$ to be known unambiguously within a certain
range (so as to account for experimental errors in measurments). We
then generate randomly-chosen pairs of observables
$(2\alpha_{eff}^i,a_i)$, and finally test whether each of these
simulated experimental ``results'' would give a value of $r^i$ that
would fit inside the range of Eq.~(\ref{eq:range}). We consider two
ranges of values for $2\alpha$: (a) $120^\circ \le 2\alpha \le
135^\circ$ and (b) $165^\circ \le 2\alpha \le 180^\circ$. We also
consider the case where the full Dalitz-plot analysis is not
available, but only $\sin 2\alpha$ is measured. In this case,
$2\alpha$ is known up to a two-fold ambiguity. Our results are shown
in Fig.~\ref{fig:results}. The darkened regions in
$(2\alpha_{eff}^i,a_i)$ parameter-space are consistent with the SM
prediction, so that NP is present everywhere in the white regions.
Should $2\alpha$ be known unambiguously, only the right-hand dark
region remains, leaving even more space for NP.

We therefore see that the analysis of $\bropi$ decays can be used to
detect NP, should it be present. Moreover, only the $\pi^+ \pi^-
\pi^0$ final state is necessary to obtain $2\alpha$ -- it is not
necessary to consider final states with two neutral pions. This is an
advantage compared to the detection of NP using $\bpipi$ decays, since
the full isospin analysis does require the two-$\pi^0$ final state.
In addition, since the $P/T$ ratios are expected to be smaller in
$\bropi$ than in $\bpipi$, the SM-consistent regions are also smaller
in the $\bropi$ channel than in the $\bpipi$ channel.

\section{Conclusion}

To summarize, $\bropi$ decays can be used to detect a discrepancy
between the phase of the mixing diagram and the phase of the penguin
diagram. Should such a discrepancy be detected, it would be a clear
signal of physics beyond the Standard Model. To carry out this method,
it is necessary to make a conservative assumption about some hadronic
parameters. QCD factorization provides us with a prediction of
$|P^i/T^i|$, $i=+-,-+$. The region in $(2\alpha_{eff}^i,a_i)$
parameter space consistent with this prediction is relatively small.
$\bropi$ is therefore an particularly useful decay channel to use for
the detection of new physics.

\section*{Acknowledgments}

I thank D. London for help throughout this work. This work was
financially supported by NSERC of Canada.


\begin{thebibliography}{0}

\bibitem{londonmrst} See D. London, these proceedings,
arXiv:hep-ph/0405241.

\bibitem{isospin} M. Gronau and D. London, {\it Phys.\ Rev.\ Lett.}
{\bf 65}, 3381 (1990).

\bibitem{charles} J. Charles, {\it Phys.\ Rev.} {\bf D59}, 054007
(1999).

\bibitem{lss01} D. London, N. Sinha and R. Sinha {\it Phys.\ Rev.}
{\bf D63}, 054015 (2001).

\bibitem{lss99} D. London, N. Sinha and R. Sinha {\it Phys.\ Rev.}
{\bf D60}, 074020 (1999).

\bibitem{np1} C.O. Dib, D. London and Y. Nir, {\it Int.\ J. Mod.\
Phys.\ A} {\bf 6}, 1253 (1991).

\bibitem{np2} Y. Grossman and M.P. Worah, {\it Phys.\ Lett.} {\bf B
395}, 241 (1997).

\bibitem{dalitz} A.E. Snyder and H.R. Quinn {\it Phys.\ Rev.} {\bf
D48}, 2139 (1993).

\bibitem{ropi} V. Pag\'e and D. London, to be published in Phys. Rev. D.

\bibitem{qcdf} M. Beneke and M. Neubert, {\it Nucl.\ Phys.} {\bf B675},
333 (2003).

\end{thebibliography}
\end{document}